\documentclass[aps,prl,twocolumn,superscriptaddress,footinbib,preprintnumbers,floatfix,showpacs]{revtex4}
\usepackage{graphicx}
\usepackage{dcolumn}
\usepackage{bm}

\newcommand{\be}{\begin{equation}}
\newcommand{\ee}{\end{equation}}
\newcommand{\ba}{\begin{eqnarray}}
\newcommand{\ea}{\end{eqnarray}}

\newcommand{\err}{\end{array}}
\newcommand{\bc}{\begin{center}}
\newcommand{\ec}{\end{center}}

\newcommand{\eg}{e.g.,~}
\newcommand{\ie}{i.e.,~}

\begin{document}
\preprint{LA-UR-07-7851}

\title[Percolation]{Origin of the Cosmic Network: Nature vs Nurture} 
\author{Sergei Shandarin}
\affiliation{Department of Physics and Astronomy, University of
  Kansas, Lawrence, KS 66045}
\author{Salman Habib}
\affiliation{T-2, Theoretical Division, Los Alamos National
  Laboratory, Los Alamos, NM 87545}
\author{Katrin Heitmann}
\affiliation{ISR-1, ISR Division, Los Alamos National
  Laboratory, Los Alamos, NM 87545}

\date{\today}

\begin{abstract}

  The large-scale structure of the Universe, as traced by the
  distribution of galaxies, is now being revealed by large-volume
  cosmological surveys. The structure is characterized by galaxies
  distributed along filaments, the filaments connecting in turn to
  form a percolating network. Our objective here is to quantitatively
  specify the underlying mechanisms that drive the formation of the
  cosmic network: By combining percolation-based analyses with N-body
  simulations of gravitational structure formation, we elucidate how
  the network has its origin in the properties of the initial density
  field ({\em nature}) and how its contrast is then amplified by the
  nonlinear mapping induced by the gravitational
  instability ({\em nurture}).

\end{abstract}

\pacs{98.65.-r, 98.65.Dx, 98.80.-k, 98.80.Bp}

\maketitle

\section{Introduction}
\label{sec:intro}

\begin{figure}    
\includegraphics[scale=0.5]{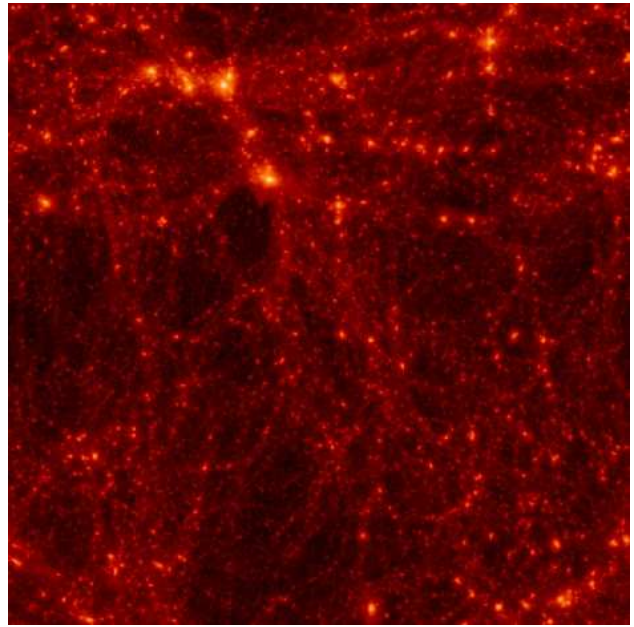} 
\caption{\label{fig_filaments} Demonstration of filamentary structure
  in N-body simulations. The projected density field is shown for a
  $\Lambda$CDM cosmology, with a box size of 64~$h^{-1}$Mpc.}
\end{figure}

Observations of the large-scale distribution of galaxies have been
underway for several decades~\cite{cfa}. Going well beyond early
``slice'' views, recent observations reveal a nontrivial
three-dimensional structure~\cite{sdss,2df}. Two key features of the
structure are immediately apparent: (i) a considerable departure from
local isotropy towards filament-like concentrations of galaxies and
(ii) a tendency of the filaments to connect into a single percolating
network spanning the entire region of observation. (A nice
visualization of large scale structure as observed by the Sloan
Digital Sky Survey can be found in Ref.~\cite{subbarao}.) A very
similar picture also emerges from cosmological N-body simulations of
the gravitational instability, the root cause of structure formation
in the cosmological standard model (Fig.~\ref{fig_filaments}).
 
Many statistics have been suggested and used to characterize
cosmological structures. The galaxy two-point correlation function and
its Fourier analog, the power spectrum, have been used extensively
since the late 1960s~\cite{tot-kih-69,pee-80}. As a way of studying
the amplitude of density fluctuations, these two-point statistics are
observationally robust and relatively straightforward to measure (in
principle), as well as to predict theoretically (see, e.g.,
Ref.~\cite{heitmann08}, and citations therein). However, being
insensitive to phases, they do not contain shape information and
cannot be used to probe the overall geometry and topology of the
large-scale structure. The full hierarchy of $n$-point correlation
functions, or the corresponding $n$-spectra, does carry complete
information about the spatial distribution of galaxies. But $n$-point
functions are difficult to measure as well as to predict. Furthermore,
the desired information can be spread in a highly nontrivial manner
over the space of $n$-point correlation functions. For these reasons,
the use of higher-point statistics has been mainly restricted to the
perturbative regime ($\langle(\delta \rho_R/\bar{\rho})^2\rangle^{1/2}
\lesssim 1$) corresponding to smoothed fields with a relatively large
smoothing scale, $R \gtrsim 5~h^{-1}$Mpc ($\bar{\rho}$ is the mean
density and $\delta\rho_R$, the density fluctuation smoothed on a
scale $R$ as defined in Eqn.~(\ref{deltaR}) below). Because of the
lack of a single obvious way to proceed (and depending on the
particular application in mind) many different statistics have been
suggested for analyzing large-scale structure observations and
simulations. These include counts in cells, two- and higher order
correlation functions~\cite{pee-80}, the Euler characteristic or
equivalently the genus curve~\cite{got-etal-86}, global and partial
Minkowski functionals~\cite{mink-func}, the void probability
function~\cite{whi-79}, minimal spanning tree~\cite{bar-etal-85}, and
many others.

In this work, we use percolation statistics to directly address the
issue of connectedness of superclusters of galaxies resulting in the
emergence of the cosmic web. Our approach is attractive for two
reasons: (i) The fact that simple scaling relations can be used to
describe the percolation properties of cosmic density fields and (ii)
that -- as we show below -- particle methods can also be conveniently
folded into the analysis.

The existence of the cosmic web in cosmological N-body simulations is
certainly manifest (Fig.~\ref{fig_filaments}). In search of a more analytic
understanding of the dynamical origin of the network topology, most
theoretical approaches start by invoking an idea due to
Zel'dovich~\cite{zeldo1,zeldo2}: approximate the full evolution by a
mapping from initial Lagrangian (${\bf q}$) coordinates to the final
Eulerian (${\bf x}$) state, via 
\be 
{\bf x}({\bf q},t)={\bf q}+D(t){\bf s}_{R}({\bf q}).
\label{eq:zeld-appr}
\ee 
Here ${\bf x}$ are the comoving coordinates, related to physical
coordinates $\bf r$ as ${\bf x}={\bf r}/a(t)$, where $a(t)$ is the
scale factor describing the uniform expansion of the universe. The
vector field ${\bf s}_{R}({\bf q})$ is determined by the density
fluctuations smoothed on a scale $R$ at the initial time $t_i$,
\be
\delta_R({\bf q},t_i) \equiv (\rho_R -\bar{\rho})/\bar{\rho} =
-D(t_i) \nabla {\bf s}_{R}
\label{deltaR}
\ee 
and the monotonically growing function $D(t)$ describes the growth of
density fluctuations in the linear regime. Additional information
present in the initial conditions may also be exploited, e.g., by
employing the deformation tensor $d_{ik} = -\partial s_i/\partial q_k$
and higher-order tensorial derivatives~\cite{sing}.

The dynamical mapping (\ref{eq:zeld-appr}) is single-valued during the
early (linear) phase of the evolution, becoming multi-valued after
orbit-crossing (nonlinear phase). The initial cosmic density field is
a realization of a Gaussian random process, specified completely by
the associated power spectrum, $P(k)$. The choice of cosmological
model fixes $a(t)$, $D(t)$, and $P(k)$. We choose to set the smoothing
of the initial perturbations to the present scale of nonlinearity, $R
\approx R_{nl}$, such that the linear {\em rms} density fluctuation
$\sigma(R_{nl}) \equiv \langle\delta_{R_{nl}}^2\rangle^{1/2} =1$ at
the present epoch. This choice of the smoothing scale is based on the
observation (made in N-body simulations) that the large-scale density
field is mainly determined by the linear power spectrum on the scales
that have just become nonlinear, \ie with $k \lesssim
R_{nl}^{-1}$~\cite{lit-etal-91}. It has also been shown that
eliminating the perturbations on scales $k \gtrsim R_{nl}^{-1}$ by
truncation of the initial spectrum~\cite{col-etal-93} or by smoothing
with a Gaussian filter~\cite{mel-etal-94} removes structures at small
scales but does not have much of an effect at large scales.

One line of attack is to consider the evolution in a given realization
of the initial conditions as a deterministic map, and derive the
properties of the network by studying generic singularities formed
during nonlinear evolution~\cite{zeldo2,sing} using the deformation
tensor, as well as higher order tensorial derivatives. Alternatively,
one can employ a probabilistic approach, investigating network
properties as determined by conditional multi-point correlation
functions between certain parameters (e.g., the shear tensor) related
to the deformation tensor at linear density peaks~\cite{cweb}. These
approaches have led to valuable insights, yet neither address the
global properties of the cosmic network directly. Both focus primarily
on the structural building blocks (halos, filaments, pancakes) and may
be loosely characterized as ``local'', even though the characteristic
scales of interest may reach tens of Mpc.

In this paper, we take up the idea of identifying the cosmic network
with percolating regions in the density field at the present time
~\cite{perc1,perc2}, and apply percolation statistics as a
quantitative global measure of the network structure~\cite{percstat}
at both early and late times. In cosmological applications of
percolation, one studies the properties of overdense and underdense
excursion sets (ES), i.e., regions where the density is greater than
some value ($\rho > \rho_c$) or less ($\rho < \rho_c$). Of particular
interest are the connectivity properties of the excursion sets as the
threshold $\rho_c$ is varied.

Previous studies of percolation regions in the initial Gaussian and
nonlinear fields -- for two spatial dimensions -- assumed the dynamics
described by the Zel'dovich approximation and showed that the
percolation region identified at the initial stage in the linear field
became a part of the percolation region identified at the nonlinear
stage~\cite{sh-zel-84}. The major reason for this conclusion was the
continuity of the mapping described by the Zel'dovich approximation.
Here, we investigate the relationship between percolating regions in
the initial (linear, Gaussian) and late-time evolved (nonlinear,
non-Gaussian) density fields using N-body simulations of a
$\Lambda$CDM cosmology. Our results establish the close -- although
not perfect -- connection between the percolating regions in the
linear and nonlinear stages of the evolution, as well as provide an
understanding of the network structure based on percolation concepts.

Studies of the percolation properties of density fields obtained in
cosmological N-body simulations with scale-free initial power spectra
($P(k) \propto k^n$) both in linear and nonlinear regimes have been
conducted previously~\cite{yess-sh-96,colombi}. In
Ref.~\cite{yess-sh-96} a suite of N-body simulations in the
Einstein-de Sitter universe (with $n= -2, -1, 0, 1$) was used to
measure the percolation transition in the mass density field at four
stages of dynamical evolution. It was shown that percolation
transitions in overdense and underdense excursion sets experience
shifts in opposite directions from that of the initial Gaussian state
(where they are obviously identical due to the symmetry of Gaussian
fields with respect to the sign of the field). In general, the
percolation transition in the overdense phase occurs at lower than the
Gaussian values while in the underdense phase it takes place at higher
values of the volume fraction (by volume fraction we mean the fraction
of the total volume in the corresponding phase). The dynamical
evolution of the density field results in shrinking the percolating
region in the overdense phase. At the same time the percolating region
in the underdense phase gains considerably in volume. It is important
to keep in mind that these are two independent statements since the
overdense and underdense percolating regions have different boundaries
in three-dimensional space. A statistically significant trend was
found relating the spectral index $n$ to percolation: the greater the
spectral index, the greater the delay in percolation as measured by
larger values of $f_{\rm p} = (V_{\rm ES}/V_{\rm tot})_{\rm
  at~percolation}$, or a correspondingly lower density threshold for
percolation, $\delta_{\rm p}/\sigma$. These results were in
qualitative agreement with Ref.~\cite{colombi} which used a set of
larger N-body simulations ($n= -2, -1, 0$). In the latter work, the
authors investigated the topology of the overdense and underdense
excursion sets at the corresponding percolation thresholds in the
density fields smoothed at various scales. They also measured
percolation thresholds in the initial Gaussian fields and found the
thresholds to be different for density fields with different power
spectra.

Our work differs from previous studies in several respects. First, we
focus on the observationally realistic case of the $\Lambda$CDM model
where the linear power spectrum cannot be described by a single power
law (previous studies of percolation in specific cosmological models
do exist, \eg Ref.~\cite{kly-sh-93}, however those models have since
been ruled out observationally). Second, our simulation volume (both
in terms of size and number of realizations) is significantly larger
than considered previously. This helps to control finite size effects
and reduces the statistical errors significantly. Finally, and most
importantly, we consider a completely new {\em dynamical} problem --
how does the (initial) linear percolating cluster map to the (final)
nonlinear percolating cluster? By quantifying the accuracy of the
mapping, we are able to distinguish between two key aspects that
underlie the structure of the cosmic web -- an already present
conspicuousness of the percolating region in the initial Gaussian
field, which we refer to as {\em nature}, and the enhancement due to
the gravitational instability, which we term {\em nurture}.

The rest of the paper is organized as follows: we first describe the
cosmological models used in the study and then present results from
percolation analysis of linear (Gaussian) density fields. After this
we discuss the results of the percolation analysis of the evolved
nonlinear (non-Guassian) density fields (filtered with the same window
as the linear density fields). In the last step, we study the accuracy
of the mapping of the linear percolating cluster into the nonlinear
percolating cluster, and end with a summary of the main results.

\section{Percolation Transition}
\subsection{Cosmological Model}

The $\Lambda$CDM cosmological model adopted here is specified by the
following parameters: the dimensionless Hubble constant, $h= H_0$/(100
km/s/Mpc) = 0.7, dimensionless dark matter and baryon densities
$\Omega_{dm}$ =0.259 and $\Omega_b=0.02~h^{-2}$, primordial spectral
index $n=1$, and a $P(k)$ normalization set by $\sigma_8
=\sigma(R_{TH}=8 h^{-1}{\rm Mpc})=0.84$, where $R_{TH}$ is the radius
of the top-hat filter -- the conventional measure of the amplitude of
the initial density fluctuations in cosmology. Our purpose is not to
consider a model that is precisely the one given by current
observations (which in any case would be a moving target), but is
nonetheless close enough to have the same generic properties.

Within this cosmology, we used two ensembles differing in box size and
filtering scale: 15 realizations generated in a 341.3~$h^{-1}$Mpc box in
linear dimension, filtered at $R =1$~$h^{-1}$Mpc ($\Lambda$CDM$_1$) and 10
realizations in a 3413~$h^{-1}$Mpc box, filtered at $R=10$~$h^{-1}$Mpc
($\Lambda$CDM$_2$). The study of the nonlinear dark matter density
field in $\Lambda$CDM$_1$ was carried out by using four statistically
independent realizations of the initial conditions. (A nonlinear field
smoothed with $R=10$~$h^{-1}$Mpc approximately corresponds to the linear
field.) The nonlinear evolution is followed using an N-body
particle-mesh code~\cite{pm}, in a simulation cube of side
341.3~$h^{-1}$Mpc. The number of particles equals the number of grid
points $N_p=N_g =512^3$, in order to smoothly sample the density
field. (The force resolution is adequate given that here we are not
interested in the small-scale distribution of matter.) In addition, we
also considered two reference power-law models ($n=-2$ and $n=4$), for
each of which 10 realizations were generated. For all four ensembles,
the filtering scale was 1.5 times the linear size of a spatial grid cell. 
\begin{figure}    
\includegraphics[scale=0.7]{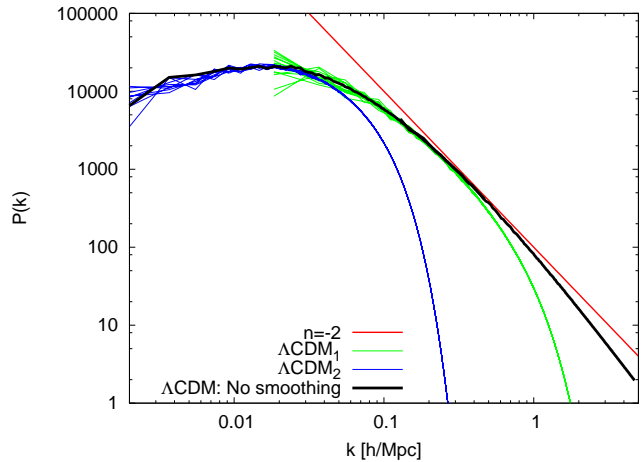} 
\caption{\label{fig:powersp}Power spectra of two $\Lambda$CDM models
 with smoothing scales, $R=1$ and 10~$h^{-1}$Mpc, and box-sizes 343.1 and
 3431~$h^{-1}$Mpc, respectively. The unsmoothed $\Lambda$CDM power
 spectrum and $P(k) \propto k^{-2}$ are also shown for reference. }
\end{figure}

\subsection{Method}

Because the initial density field so strongly controls properties of
the structure at late
times~\cite{lit-etal-91,col-etal-93,mel-etal-94}, we first discuss
percolation in Gaussian fields with different $P(k)$, corresponding to
four ensemble choices: two with $\Lambda$CDM $P(k)$ (parameters as
above) and two with power-law spectra $P(k) \propto k^n$ with $n=-2$
and $4$ (all on a $512^3$ grid). The power spectrum in the
$\Lambda$CDM model is not a simple power law. Instead, in the limit of
small $k$, $P(k) \propto k$ while in the opposite limit, $P \propto
k^{-3}\ln(k/k_c)$, and at the present scale of nonlinearity (roughly),
$P \propto k^{-1.5}$ (Fig. \ref{fig:powersp}). In the not so distant
past the scale of nonlinearity was smaller and the effective slope was
steeper, approximately $n=-2$; in principle this evolutionary stage is
observable. The choice of $n=4$ corresponds to the so called `minimal
power spectrum' on large scales~\cite{dor-zel-74} (this power spectrum
would arise from initial conditions which have power only on small
spatial scales).

Percolating sites are identified using the ``friends-of-friends''
algorithm on a cubic mesh (only the eight closest neighbors were
considered as immediate friends) for the (thresholded) density field
filtered on a scale $R$ with a Gaussian window. As the density
threshold is reduced from a high starting value, the volume in the
overdense excursion set increases monotonically until percolation
occurs, therefore the volume fraction of the excursion set itself can
be used as a proxy for the density threshold. (The same argument also
holds for the underdense excursion set.) For the linear density field,
percolation curves were averaged over overdense and underdense
excursion sets exploiting the exact statistical symmetry of Gaussian
fields. The power spectra in the unsmoothed $\Lambda$CDM model are
shown in Fig. \ref{fig:powersp} along with the power spectra in all
realizations of the $\Lambda$CDM$_1$ and $\Lambda$CDM$_2$ models; a
power-law spectrum with $P(k) \propto k^{-2}$ is also shown for
reference.


\subsection{Linear Stage}
We first focus on the initial stage of structure formation, i.e., any
epoch after decoupling of baryonic matter and radiation with only
linear density fluctuations on the scales of interest. For the chosen
resolution scales (set by the particle number and box size), an
initial redshift of $z_i=50$ comfortably satisfies this criterion.

\begin{figure}        
\includegraphics[scale=0.68]{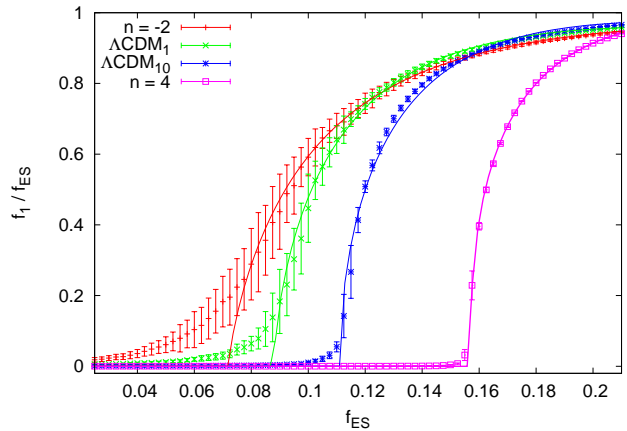} 
\caption{\label{fig:perc-z50} Percolation transitions in Gaussian
  density fields. Results for power-law spectra $P(k) \propto k^n$
  with $n=-2$ (extreme left) and $n=4$ (extreme right), $\Lambda$CDM
  -- the two interior curves -- smoothed at $R=1$ (left) and
  $R=10$~$h^{-1}$Mpc (right).  Smooth curves are fits to the percolation
  ansatz (\ref{eq:perc-ansatz}); error bars are $1 \sigma$ (statistical).}
\end{figure}

In order to carry out our percolation analysis we used the volume
fraction of the excursion set rather than the density itself as the
relevant parameter. The resulting data -- the percolation curves --
turn out to be well-described by the single-variable percolation
scaling ansatz~\cite{percans}  
\ba f_1 =A (f_{\rm ES} - f_{\rm p})^{\nu}~~~{\rm at} ~~~f_{\rm ES} >
f_{\rm p}, 
\label{eq:perc-ansatz}
\ea 
where $f_{\rm ES}$ is the volume fraction, i.e., the fraction of the
volume in the excursion set ($f_{\rm ES} \equiv V_{\rm ES}/V_{\rm
  tot}$), $f_1$, the volume fraction of the largest region, and,
$f_{\rm p}$, the value of the volume fraction when the percolation
transition occurs. In Fig.~\ref{fig:perc-z50}, the quantity,
$f_1/f_{\rm ES}$, as numerically determined, is plotted as a function
of $f_{\rm ES}$, and undergoes a sudden growth from zero to unity at
the percolation transition. As is evident, the fitting form
(\ref{eq:perc-ansatz}) provides a good fit to the data for all four
models. The parameters of the fits are given in Table~\ref{tab:z50}.
\begin{table}          
  \caption{\label{tab:z50} Values for the percolation ansatz
    (\ref{eq:perc-ansatz}) parameters ($A$ and $\nu$) for Gaussian fields, 
    the percolation density threshold $\delta_p$ corresponds to
    $f_{\rm p}$, the volume fraction at percolation.}  
\begin{ruledtabular}
\begin{tabular}{ccccc} 
\hline
Model                      & $f_{\rm p}$ & $A$    & $\nu$ &  $\delta_p/\sigma$ \\ 
\hline
n=4                          & 0.157   & 0.61 & 0.38 & 1.006 \\
$\Lambda$CDM$_2$                & 0.111   & 0.66 & 0.51 & 1.22 \\
$\Lambda$CDM$_1$                   & 0.089   & 0.75 & 0.62 & 1.36\\
 n=-2                       & 0.072   & 0.89 & 0.76 & 1.46\\
 \end{tabular}
\end{ruledtabular}
\end{table}
\begin{figure}    
\includegraphics[scale=0.7]{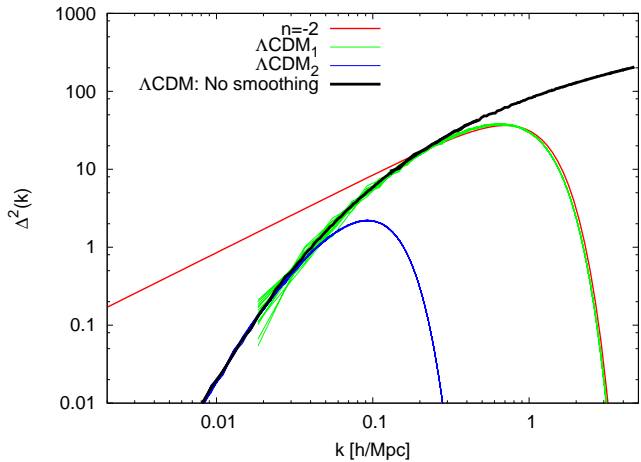} 
\caption{\label{fig:Del2} The function $\Delta^2(k) = k^3 P(k)$ plotted
  for the power spectra shown  in Fig.~\ref{fig:powersp}. The power
  law spectrum with $n=-2$ is smoothed with the same window as
  the $\Lambda$CDM$_1$ model. The normalization of $\Delta^2(k)$ is
  not important since it determines only the variance of the field.}
\end{figure}

The data shown in Fig.~\ref{fig:perc-z50} clearly demonstrate that
percolation occurs at lower volume fractions with growth of power on
large scales, in qualitative agreement with Ref.~\cite{colombi}. The
corresponding density threshold reaches its (lower) limiting value
($\delta_p = \sigma$, $f_{\rm p}=0.157$), following a conjecture due
to Ziman~\cite{ziman} who suggested that the percolation transition
should occur exactly at the $\delta=\sigma$ level, for Gaussian fields
with dominating small-scale power. At the other extreme considered
here, $P(k) \propto k^{-2}$, where the field is dominated by
large-scale power, $f_{\rm p}=0.072$, and corresponds to the highest
value of the density threshold $\delta_p/\sigma =1.46$. This is
slightly lower than $\delta_p/\sigma =1.6\pm0.1$ as found in
Ref.~\cite{colombi}. The minor disagreement may be due to how the
percolation threshold was determined in the two cases, as well as due
to finite-volume limitations. [It is also worth noting that our set of
fields includes models with nonpower law $P(k)$.] Each parameter in
Table~\ref{tab:z50} varies monotonically with increase of power on
large scales: $f_p$ decreases while all the rest increase.

From the results of Fig.~\ref{fig:perc-z50} it is clear that the
percolation transition becomes more gradual as the (effective)
spectral index $n$ decreases, deviating from the percolation scaling
ansatz at small $f_1/f_{\rm ES}$. Therefore the fit given by
Eqn.~(\ref{eq:perc-ansatz}) becomes less accurate and $f_{\rm p}$
determined by this method becomes less reliable. This suggests that
even larger volumes than the ones used here may be necessary to study
such cases. Other factors that may affect the accuracy of the
determination of the percolation threshold are discreteness, accuracy
of computing the volume of the excursion set, filtering, and the
properties of the largest cluster. As the present study is not devoted
to an accurate calculation of percolation thresholds in Gaussian
fields, we leave these questions to future work.

For scale-free power spectra, $P(k) \propto k^n$, the dependence of
the percolation transition in Gaussian fields on the power spectrum
was observed and qualitatively discussed in Ref.~\cite{yess-sh-96} and
then quantified in Ref.~\cite{colombi}. Here, along with two types of
scale-free fields, we also study the Gaussian field corresponding to
the $\Lambda$CDM model smoothed at two different scales
(Fig.~\ref{fig:powersp}). As mentioned earlier the large-scale
structure at the nonlinear stage is mainly determined by the initial
power spectrum on scales with $k \lesssim
R_{nl}^{-1}$~\cite{lit-etal-91,col-etal-93,mel-etal-94} and therefore
can be approximately characterized by the effective slope at this
scale. It is thus interesting to investigate how well the effective
slope at the scales contributing most to the variance of the field
determines the percolation transition in the linear regime.

As a first candidate, we study $\Delta^2(k) \equiv k^3 P(k)$, which
represents the contribution of power to the variance of the density
field, per unit logarithmic interval in $k$. In particular, the
position of the maximum of $\Delta^2(k)$ corresponds to the wavelength
with maximum contribution to the variance of the field.
Figure~\ref{fig:Del2} shows that the maximum contribution arises from
$k \approx 0.65 $ and $0.09$~$h$Mpc$^{-1}$ for the $\Lambda$CDM$_1$
and $\Lambda$CDM$_2$ models respectively. The initial \ie unsmoothed
spectrum has the corresponding effective slopes $n_{\rm eff} = d \log
P/d\log k$ $\approx -2.2$ and $-1.4$ at these wavenumbers.

Referring to Table~\ref{tab:z50}, two subsets of the models
represented there -- one subset consisting of the power law model with
$n=-2$ , $\Lambda$CDM$_2$ with $n_{\rm eff}=-1.4$, and the power law
model with $n=4$, the other consisting of $\Lambda$CDM$_1$ with
$n_{\rm eff}=-2.2$, $\Lambda$CDM$_2$ with $n_{\rm eff}=-1.4$, and the
power law model with $n=4$, both demonstrate monotonic growth of the
percolation volume fraction with increase of effective slope: $f_p$ =
0.07, 0.11, 0.16 and $f_p$ = 0.09, 0.11, 016 respectively. However,
percolation in the $\Lambda$CDM$_1$ model with $n_{\rm eff} = -2.2$
formally takes place at a higher value of the volume fraction, $f_p$ =
0.09, than for the power law model with $n = -2$, where $f_p= 0.07$.
Regardless of the limited accuracy inherent in measuring the values of
$f_p$ from the percolation ansatz~(\ref{eq:perc-ansatz}), it does
appear reliable to conclude directly from the data points of
Fig.~\ref{fig:perc-z50} that the percolation transition in the power
law model with $n = -2$ occurs at a lower volume fraction then in the
$\Lambda$CDM$_1$ model. Additionally, the data in
Fig.~\ref{fig:powersp} show that the effective slope of the unsmoothed
$\Lambda$CDM model is slightly more negative than $n = -2$ at $k
\approx 0.9$~$h$Mpc$^{-1}$ (where $\Delta^2(k)$ peaks for the $n=-2$
power spectrum) .


An alternative possibility is to use the power per equal interval in
$k$ rather than in $\log k$, \ie $E(k) = k^2 P(k)$ instead of
$\Delta^2(k) = k^3 P(k)$; the former choice being widely used in the
theory of turbulence~\cite{mon-yag-75}. Again, by finding the maximum
of $E(k)$ and then the effective slope in the unsmoothed spectra at
these points, one finds that all four models listed in ascending order
of $f_p$ (\ie $n = -2$, $\Lambda$CDM$_1$, $\Lambda$CDM$_1$, and $n =
4$) correspond to monotonically increasing effective slopes, $n_{\rm
  eff}$ = -2, -1.97, -1.1, and 4. The difference in $n_{\rm eff}$
between the power law model with $n = -2$ and the $\Lambda$CDM$_1$
model is obviously marginal, but the effective slopes at the maxima of
$E(k)$ do appear to correspond better to the volume fractions at
percolation. The maxima of $E(k)$ are obviously shifted to smaller $k$
with respect to the maxima of $\Delta^2(k)$, therefore $E(k)$ takes
into account power on slightly smaller scales than the scales of the
maxima of $\Delta^2(k)$. A more detailed evaluation of this issue is a
topic for further study and beyond the scope of the current work.

Despite the qualifications in the above discussion, the $\Lambda$CDM
percolation curves do display trends qualitatively similar to those
seen in the bracketing power-law cases. Filtering the density field at
smaller scales corresponds to a greater negative effective spectral
slope than that due to filtering at larger scales
(Fig.~\ref{fig:powersp}), the effect of the moving cutoff in $k$-space
leading therefore to a higher density threshold and lower volume
fraction at percolation, as demonstrated by the two interior curves in
Fig.~\ref{fig:perc-z50}. In particular, this means that the structure
formed earlier at greater redshifts was formed from the Gaussian field
percolating better than the field corresponding to the current
structure.

The dependence of the percolation transition parameters on the power
spectrum of a Gaussian field shows that such fields can differ not
only locally (\eg in the shape of peaks, etc.~\cite{bar-etal-86}) but
also globally. The higher percolation threshold, $\delta_p$, or
equivalently lower value of $f_{\rm p}$, indicates that Gaussian
fields with relatively more power on large scales have a greater
degree of connectedness. Therefore, the relatively large negative
slope of the linear power spectrum at the scale of nonlinearity in the
$\Lambda$CDM cosmology is an important factor determining the origin
of the cosmic web. Because this feature is already present in the
initial conditions, later to be amplified by the gravitational
instability, we refer to it as the `nature' factor influencing the
formation of the cosmic web.

\subsection{Nonlinear Stage}

\begin{figure}   
\includegraphics[scale=0.68]{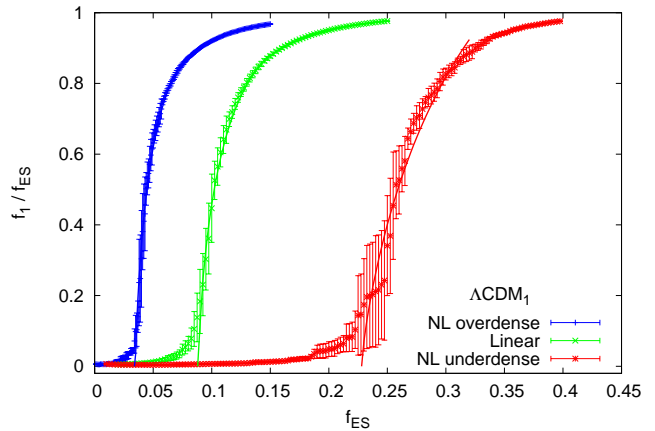}  
\caption{\label{fig:perc-z0} $\Lambda$CDM$_1$ percolation transitions
  (filter scale set to $R=1$~$h^{-1}$Mpc). Nonlinear stage: overdensity (blue/left) and
  underdensity (red/right) excursion sets at $z=0$; linear (Gaussian) stage
  (green/middle) at $z=50$. Solid curves are percolation ansatz fits (cf.
  Eqn.~\ref{eq:perc-ansatz} and Table~\ref{tab:z0}).}
\end{figure}

Next we turn to a percolation transition analysis of the nonlinear,
and therefore non-Gaussian, density field at $z=0$ for the
$\Lambda$CDM$_1$ model. As shown in Fig.~\ref{fig:perc-z0}, the
percolation threshold for the overdense excursion set at the nonlinear
stage (as measured by the volume fraction) is remarkably lower than
that for the initial Gaussian field: $f_{\rm p}(z=0) = 0.035$ while
$f_{\rm p}(z=50) = 0.089$. This result is as expected from the
Zel'dovich approximation~\cite{sh-zel-84}. The evolution of the
density field results roughly in compression of overdense regions and
expansion of underdense regions. This qualitatively correct feature is
complicated by the fragmentation of the dense regions into
gravitationally bound halos that might make percolation more
difficult. However, the relative strength of both effects strongly
depends on the power spectrum of the initial Gaussian field. This
study suggests that despite the strong fragmentation of the linear
percolating region, the overall spatial distribution of the fragments
markedly preserves the topology of the linear percolating region, at
least in the important case of $\Lambda$CDM.

Compared to the linear case, the density threshold for
percolation in the nonlinear regime is significantly larger. (The mean
values and standard deviations are obtained from four realizations of
the nonlinear stage and thirty realizations of the linear Gaussian
field.) A visual comparison can be made by comparing the nonlinear and
linear percolating clusters. These are shown in
Figs.~\ref{fig:slab-z50} (the blue region is the linear percolating
cluster) and \ref{fig:slab-z0} (the yellow region is the nonlinear
percolating cluster). At the same time the total mass in the nonlinear
percolating cluster is about 15 times greater than that in the
corresponding linear cluster: the mass fraction $M_1\approx 0.25$ at
$z=0$, while $M_1\approx 0.018$ at $z=50$. The rightmost curve in
Fig.~\ref{fig:perc-z0} shows that the percolation transition in the
underdense excursion set occurs at considerably greater volume
fraction than in the initial Gaussian field: $f_{\rm p}(z=0) = 0.23$
while $f_{\rm p}(z=50) = 0.089$. 
\begin{table}   
  \caption{\label{tab:z0} Values for the percolation ansatz
    (Eqn. \ref{eq:perc-ansatz}) parameters for 
    underdense and overdense excursion sets at the 
    nonlinear stage of evolution, the percolation density threshold $\delta_p$
    corresponds to $f_{\rm p}$.}  
\begin{ruledtabular}
\begin{tabular}{ccccc} 
\hline
Model                      & $f_{\rm p}$ & $A$    & $\nu$ &  $\delta_p$ \\ 
\hline
NL underdense       & 0.228    & 1.80 &  0.76 &  -0.80\\
 NL overdense        & 0.035   & 0.73 &0.75 & 3.31\\
 \end{tabular}
\end{ruledtabular}
\end{table}

The parameters of the fitting curves are given in Table~\ref{tab:z0}.
It is worth noting that the percolation transitions in all four tested
Gaussian fields are significantly greater than the percolation
threshold -- measured by volume fraction -- for the over-dense phase
but smaller than that for the under-dense phase of the nonlinear
density field generated by the gravitational instability. Expanding
the study of percolation to $n \rightarrow -3$ is relatively
difficult, therefore whether the above result holds in this limit
remains to be seen. It is also worth pointing out that the values of
the exponent $\nu$ are similar for both overdense and underdense
excursion sets, approximately coinciding with the value of $\nu$
obtained for the scale-free Gaussian field with $n = -2$. The
amplitude $A$ is different in all three cases, however.

\section{Mapping Accuracy}

\begin{figure*}   
\includegraphics[scale=0.58]{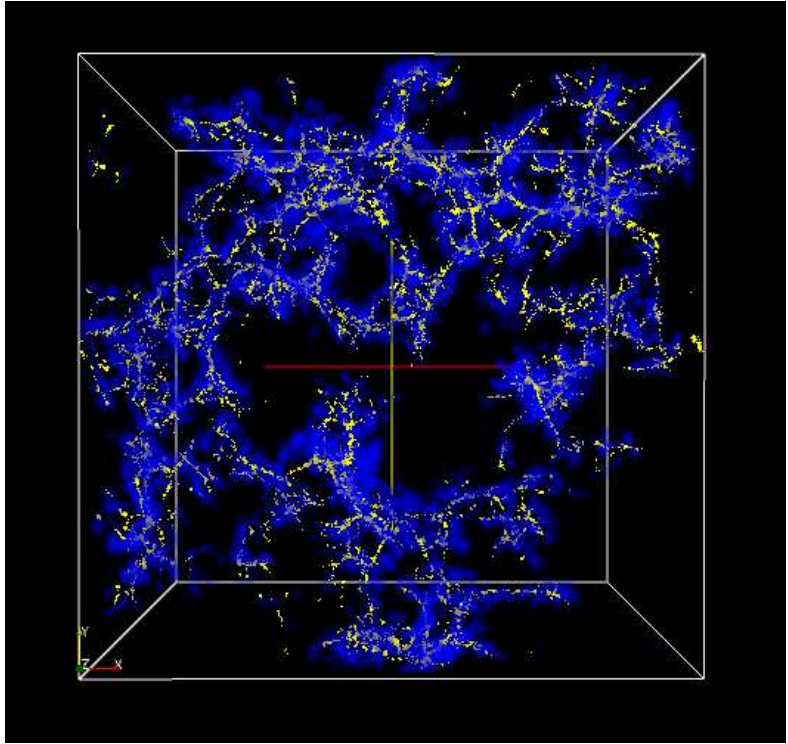}
\caption{\label{fig:slab-z50} Forward mapping of the linear
  percolating cluster. The mapping takes the percolating region
  identified in the linear regime at $z=50$ (blue) to the final
  forward-mapped region at $z=0$ (yellow, see text). Note
 the fragmentation of the initial percolating region after the
 mapping. For visual clarity, a 70~$h^{-1}$Mpc thick slab cut out of
 the full simulation cube is shown. } 
\end{figure*}
\begin{figure*}   
\includegraphics[scale=0.58]{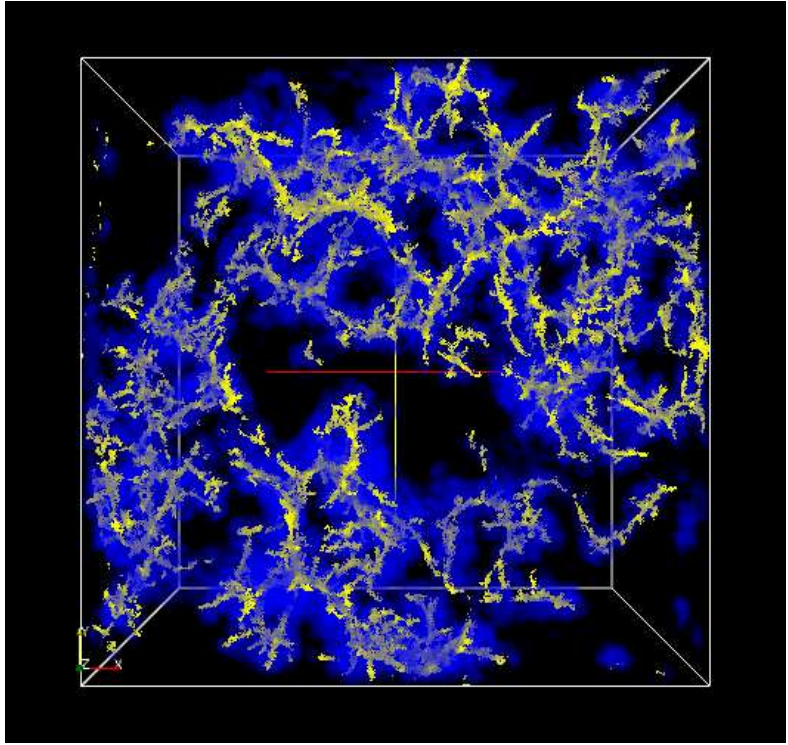}
\caption{\label{fig:slab-z0} Backward mapping of the percolating
 region identified in the nonlinear regime at $z=0$ (yellow) to the
 initial stage at $z=50$ (blue). In contrast to the case of the
 forward map, the backwards map takes a percolating region into
 another percolating region. The slab is the same as in  Fig.
 \ref{fig:slab-z50}.  } 
\end{figure*}

Standard Eulerian perturbation theory at linear order does not move
particles. However, if one takes the percolating cluster in the linear
density field, smooths it at the scale of nonlinearity, and maps it
using the Zel'dovich approximation then in effect one has a (linear
theory) prediction for the nonlinear percolating cluster. Intuitively,
one may expect that the percolating cluster in the linear field should
approximately map into the exact nonlinear percolating cluster, but
the accuracy of this proposition has not been directly tested. In this
section, we focus on whether the linear percolating cluster indeed
maps onto the nonlinear percolating cluster.

We begin by considering the mapping from the initial, linear stage of
the evolution ($z=50$), to the final nonlinear stage ($z=0$), the {\em
  forward map}. This requires two additional steps. First, we choose a
fiducial value for the density threshold, $\delta^{(f)}$, to carry out
the percolation analysis. We set this to correspond to the point where
the largest percolating region occupies 50\% of the excursion set
volume, i.e., $f^{(f)}_1/f_{\rm ES} = 0.5$. With this choice, the
percolation region is well developed, yet not completely dominating
the excursion set. This definition is different from that used in the
previous analysis (which followed the conventional practice in
percolation theory). The reason for this change is dictated by the
fact that at the (conventional) percolation transition, the
percolating cluster occupies an infinitesimal volume ($f_1/f_{\rm ES}
\rightarrow 0$). This means that in a finite box its volume is much
smaller than the volume of the box and is thus subject to very large
fluctuations. Our choice makes the definition of the percolating
cluster more robust and less subject to the precise ``turn-on'' of
percolation (Cf. Figs.~\ref{fig:perc-z50} and \ref{fig:perc-z0}).

The second step involves multiple back and forth switching from
density fields defined on an invariant spatial grid to particles
subject to dynamic motion. Although the percolation analysis can
be carried out on a set of points it is numerically more efficient 
on a regular grid.  In this study we used a very fast algorithm
that detects the percolation transition by performing the cluster
analysis on the grid \cite{percans}. In particular, it defines the percolating
cluster as a set of grid sites. In order to study the forward mapping
of the initial percolating cluster (identified as a set of grid
sites), we first need to determine the set of particles that are
associated with it. Following the motion of these selected particles
will allow us to study how the initial percolating region maps into
the nonlinear stage.

As a definition, the initial percolating set is taken to be all those
particles lying within a distance smaller than half a grid unit (along
all three orthogonal axes) from any grid site that belongs to the
initial percolating region. After mapping particles to the final
nonlinear state we find all the sites on the grid that satisfy the
very same criterion. The set of these new sites at $z=0$ approximates
the map of the initial percolating cluster and can be analyzed by the
same fast method. We stress that this forward mapping of the linear
percolating cluster comprises only a relatively small fraction of
particles in the nonlinear percolating cluster because a fundamental
feature of the nonlinear evolution is the crossing of orbits and the
resulting formation of multi-stream flow regions. The forward mapping
of the percolating cluster found in the linear density field forms
only a fraction of those streams, and many others come from regions
beyond it. In a similar manner, by starting with the nonlinear
percolating cluster at $z=0$, we introduce a {\it backward} mapping
that finds {\em all} sites at $z=50$ that were mapped to $z=0$ and
formed the nonlinear percolating cluster.

With the particle-based definition of percolation applied to the
$\Lambda$CDM$_1$ cosmology, we find that the particles in the initial
percolating set make up 4.6\% of the total number of particles in the
excursion set -- quite close to the volume fraction of the percolating
region as defined on the grid (0.045). The small excess is due to the
higher than mean density in the percolating region: the two
definitions of a percolating region (on grid and by particles) are not
identical. But we do expect them to be very similar during the linear
stage of the evolution. To verify this expectation we go back from the
selected particles to the grid by identifying every grid site closer
than half a grid unit to any particle from the percolating set. We
then perform a cluster analysis on these sites. Ideally we would find
exactly the same grid sites as in the initial percolating cluster. The
actual results show good but not perfect correspondence between the
two procedures.

Figure \ref{fig:slab-z50} illustrates the forward mapping of the
initial percolating set (blue) into the nonlinear stage (yellow). Both
structures are displayed in comoving coordinates to scale out the
uniform Hubble expansion. To avoid excessive projections, every tenth
particle is shown in a 70~$h^{-1}$Mpc thick slab cut out of the simulation
cube. The mapping obviously results in a large reduction of the volume
as expected, therefore the nonlinear (yellow) cluster is much smaller
by volume than the corresponding linear cluster (blue). A closer look
also suggests that the initially connected structure at $z=50$ becomes
fragmented after nonlinear mapping to $z=0$. We discuss this in more
detail below.

\begin{figure}                     
\includegraphics[scale=0.68]{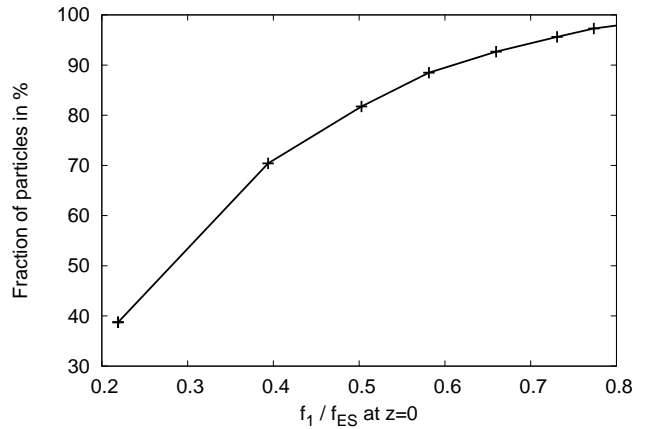}  
\caption{\label{fig:frac-z50-in-z0} Fraction of the particles from the
  initial percolating set ($z=50$) in the percolating region at $z=0$
  as a function of $f_1/f_{\rm ES}$ at $z=0$. }
\end{figure}

In the first step of the analysis, we compute the fraction of
particles from the initial percolating set that end up in the final
percolating region, as defined on the grid at various thresholds, and
shown for the fiducial $\delta^{(f)} = 2.76$ in Fig.~\ref{fig:slab-z0} in
yellow. (We determine whether a particle belongs to the percolating
region at $z=0$ using the same criterion of proximity of particles to
grid sites as above, filtering the nonlinear density field also at
$R$=1~$h^{-1}$Mpc.) The fraction of particles from the initial
percolating set transported into the final percolating region is
plotted as a function of the ratio $f_1/f_{\rm ES}(z=0)$,
characterizing the size of the percolating cluster at $z=0$, in
Fig.~\ref{fig:frac-z50-in-z0}. This fraction grows monotonically from
about 40\% at $f_1/f_{\rm
  ES}(z=0) \approx 0.22$ to more than 95\% at $f_1/f_{\rm ES}(z=0)
\approx 0.8$ reaching about 80\% at the fiducial value $f^{(f)}_1/f_{\rm
  ES}(z=0) = 0.5$. The point of this result is that the progenitor of
the ``cosmic web'' defined as the percolating cluster of the initial
Gaussian density field, is a fair, albeit not perfect, ``backbone'' of
the web at a later, nonlinearly evolved stage.

It is also of interest to investigate the percolation properties of
the initial set of percolating particles at the final stage of
evolution. Using the criterion described above, we generated the set
of grid sites at $z=0$ close to the selected particles (those
representing the percolating cluster at $z=50$) after they had moved
to their final positions at $z=0$. We find that these sites comprise
less than 0.003 of the total volume. This should be compared with
their initial volume occupation fraction of 0.045 at $z=50$, resulting
in a compression factor of about 15 on average. This compression
reflects the fact that overdense regions tend to collapse and
therefore more than one particle becomes associated by the proximity
criterion to the same site at the nonlinear stage. This is clearly
seen in Fig.~\ref{fig:slab-z50}. By $z=0$, the initially connected
region fragments into large numbers ($\sim 2.5\times10^4$) of isolated
regions and none of them percolates. The largest such region has a
volume of $\sim$2000~$h^{-3}$Mpc$^3$. Although this volume is quite
large by astronomical standards (albeit much smaller than that set by
the scale of homogeneity which has a linear dimension of
$\sim$70~$h^{-1}$Mpc), it comprises only 0.017 of the volume occupied
by the particles from the initial percolating set after they are
mapped to $z=0$. These numbers illustrate the prolific fragmentation
of the linear percolating cluster in the course of the nonlinear
mapping.

We now take the percolating region at $z=0$, shown in
Fig.~\ref{fig:slab-z0} in yellow, back to the linear stage -- the {\em
  backward map} (blue region in Fig.~\ref{fig:slab-z0} with every
twentieth particle plotted). As an example, we select the percolating
region built from the smoothed density field at $\delta_{f} = 2.76$
corresponding to $f^{(f)}_1/f_{\rm ES}(z=0) = 0.5$.  This occupies $8.4
\times 10^{5}$~$h^{-3}$Mpc$^3$ or 0.02 of all grid sites or equivalently
about 0.02 of the volume of the box.  The remaining half of the
excursion set consists of $\sim 2\times 10^4$ isolated regions.  It is
worth stressing that approximating the percolating region at the
nonlinear stage by the particle set as described above, we find that
the two representations (particle vs. grid) differ considerably more
than they do in the linear regime. This is because the particle
representation has a significantly higher resolution than the smoothed
density field on the grid. In any case, the backward map must be
applied using the particle representation; inverse mapping the
particles from the largest percolating set back to the linear stage,
we find that they occupy about a quarter of the total volume and also
form a percolating region (in contrast to the case of the forward
map).

\section{Summary}

We now summarize our key results. We began by confirming -- with some
minor discrepancies -- earlier findings of the dependence of the
percolation threshold in Gaussian random fields on the power spectrum.
However, the purpose of our study was not the precise measurement of
the percolation thresholds for Gaussian fields but to establish the
existence of strong percolation in the initial/linear density field in
the $\Lambda$CDM cosmological model -- the {\em
  nature} factor in the formation of the cosmic web. This factor is
determined exclusively by the density (or density contrast) field,
$\delta \equiv (\rho -\bar{\rho})/\bar{\rho}$ in Eulerian linear
theory.


We found that the nonlinear mapping from the linear stage of cosmic
evolution ($z=50$) to the final nonlinear stage at the current epoch
($z=0$) results in a considerable amplification of percolation
features: the percolation volume fraction reduces from $f_p(z=50)
\approx 0.09$ to $f_p(z=0) \approx 0.035$, the half-fill volume of the
percolating cluster at the fiducial percolation volume fraction
reduces from $f^{(f)}_1(z=50)\approx 0.05$ to $f^{(f)}_1(z=0)\approx
0.02$, while its mass greatly increases from $m^{(f)}_1(z=50)\approx
0.025$ to $m^{(f)}_1(z=0) \approx 0.25$. All these parameter values
indicate that the cosmic web becomes considerably thinner and more
conspicuous than its progenitor in the linear density field. We call
this dynamical influence of the gravitational instability, the {\em
  nurture} factor in the formation of the cosmic web because it was
not present in the initial field and results entirely from the mapping
itself. Conventional linear theory does not consider mapping at all;
the Zel'dovich approximation suggests that the initial displacement
field, ${\bf s} = -\partial \Phi_{\rm lin} /\partial {\bf q}$ smoothed
at the current scale of nonlinearity determines the dominant features
of the mapping.

We conclude that the robust contrast of network structure -- so
characteristic of N-body simulations -- is determined by a combination
of two factors: the already-present conspicuousness of the percolating
region in the initial Gaussian field ({\em nature}) and the effects of
nonlinear mapping ({\em nurture}). The first is quantified by the
relatively high density threshold ($\delta^{(f)} = 1.27 \sigma$) and
low volume fraction of the excursion set ($f^{(f)} \approx 0.1$) at
the percolation transition. This is directly related to the character
of the linear power spectrum -- a relatively large negative effective
slope in the relevant range of scales from $k = 0.1$ to
$1$~$h$Mpc$^{-1}$. The second factor is measured by the further
decrease of the volume fraction ($f^{(f)} \approx 0.05$) at
percolation. The percolation threshold in the density field filtered
on $R = 1$~$h^{-1}$Mpc is $\delta^{(f)}=2.76$ with the percolating
region containing about a quarter of the total mass. We conclude that
the nurture factor arising from the dynamical mapping itself is more
important for explaining the conspicuousness of the cosmic web.

The topology of the web can also be quantified by the Euler
characteristic, but is not equivalent to percolation
statistics~\cite{sah-sat-sh-97}. In particular, the density of the
Euler characteristic, $\chi$, as a function of the level, $\nu =
\delta/\sigma_{\delta}$ has the universal shape, $n_{\chi} = N_{\chi}
(\nu^2-1)\exp(-\nu^2/2)$ for all Gaussian fields regardless of the
power spectrum~\cite{euler-ch}. The power spectrum determines only its
amplitude, $N_{\chi} =2(\langle k^2\rangle/3)^{3/2}/(2\pi)^2$, where
$\langle k\rangle^2= \int k^2 P(k) k^2 dk / \int P(k) k^2 dk
$~\cite{bar-etal-86}. Thus, the standard diagnostics used in cosmology
for characterization of the structure: ``meatball shift'', $\Delta
\nu$, ``number of voids'', $A_V$, and ``number of clusters'', $A_C$
(see \eg Ref.~\cite{gott-etal-09}) designed for studies of
non-Gaussian features of the structure are unable to detect the
`nature' factor in structure formation.

We note that the mapping of the linear percolating cluster into the
nonlinear stage is not flawless. The percolating region at the
nonlinear stage taken at the half-fill percolation volume fraction
contains about 80\% of the percolating region found in the linear
density field, also taken at the half-fill percolation volume
fraction. This is close, though not perfect, correspondence. The
half-fill linear percolating cluster makes up about 5\%, while the
nonlinear one about a quarter, of the total mass, i.e., five times
greater. The initial percolating region fragments in the course of
evolution into a large number (more than $2.5\times10^4$ in the
current simulation) of isolated clumps by $z=0$, none of which
percolates at $z=0$.


A comparison of the percolation properties of the galaxy distributions
in the $\Lambda$CDM model and galaxy redshift surveys requires mock
redshift catalogs obtained from simulations covering volumes similar
to the 2dF or SDSS surveys. An additional theoretical complication is
related to the fact that real galaxy distributions are available only
in redshift space. Redshift space is fundamentally anisotropic in the sense that 
the statistical properties of the large-scale structure are different along the line of sight
and in the orthogonal directions \cite{z-space-anis}.  This may
require additional refining of the technique discussed here for its
analysis. However, some observational studies have already been
carried out and have succeeded in measuring a significant signal of
filamentarity in redshift catalogs~\cite{perc-observ}.

SS acknowledges sabbatical support at Los Alamos National Laboratory.
SH and KH acknowledge support from LANL's LDRD and institutional open
supercomputing programs. This research was initiated at the Aspen
Center for Physics in 2005.


\begin{thebibliography}{99}
\bibitem{cfa} S.A.~Gregory and L.A.~Thompson, Astrophys.~J. {\bf 222},
  784 (1978); G.~Chincarini and H.~Rood, Astrophys.~J. {\bf 230}, 648
  (1979); V. de Lapparent, M.J.~Geller, J.P.~Huchra,
  Astrophys.~J. {\bf 302}, L1 (1986); H.~Lin et al.,
  Astrophys.~J. {\bf 464}, 60 (1996).
\bibitem{sdss} J.K.~Adelman-McCarthy et al., Astrophys. J. Supp. {\bf
    172}, 634 (2007).
\bibitem{2df} M.~Colless et al., MNRAS {\bf 328}, 1039 (2001).   
\bibitem{subbarao} M.U.~SubbaRao, M.A.~Arag\'on-Calvo, H.W.~Chen,
  J.M.~Quashnock, A.S.~Szalay, and D.G.~York, New J. Phys. {\bf 10},
  125015 (2008).     
\bibitem{tot-kih-69} H. Totsui and T. Kihara, Publ. Astron. Soc.
  Japan {\bf 21}, 221 (1969). 
\bibitem{pee-80} P.J.E.~Peebles, {\em The Large Scale Structure of the
    Universe},  Princeton Univ. Press, Princeton  (1980). 
\bibitem{heitmann08} K.~Heitmann, M.~White, C.~Wagner, S.~Habib, and
 D.~Higdon, Astrophys. J. (submitted), arXiv:0812.1052 [astro-ph].    
\bibitem{got-etal-86} J.R.~Gott III, M.~Dickinson, and A.L.~Melott,
  Astrophys.~J. {\bf 306}, 341 (1986).
\bibitem{mink-func} K.R.~Mecke, T.~Buchert, and H.~Wagner, Astron. \&
  Astrophys. {\bf 288}, 697 (1994); J.V.~Sheth, V.~Sahni,
  S.F.~Shandarin, and B.S.~Sathyaprakash, MNRAS {\bf 343}, 22 (2003).
\bibitem{whi-79} S.D.M.~White, MNRAS {\bf 186}, 145 (1979).
\bibitem{bar-etal-85} J.D.~Barrow, S.P.~Bhavsar, and D.H.~Sonda,
  MNRAS {\bf 216}, 17 (1985).                                               
\bibitem{zeldo1} Ya.B.~Zel'dovich, Astron. \& Astrophys. {\bf 5}, 84 (1970). 
\bibitem{zeldo2} S.F.~Shandarin and Ya.B.~Zel'dovich,
  Rev. Mod. Phys. {\bf 61}, 185 (1989). 
\bibitem{sing} V.I.~Arnold, S.F.~Shandarin, and Ya.B.~Zel'dovich,
 Geophys. Astrophys. Fluid Dyn. {\bf 20}, 111 (1982). 
\bibitem{lit-etal-91} B.~Little, D.H.~Weinberg  and C.~Park, MNRAS
  {\bf 253}, 295 (1991).  
\bibitem{col-etal-93} P.~Coles, A.L.~Melott, and S.F.~Shandarin,
  MNRAS {\bf 260}, 765 (1993). 
\bibitem{mel-etal-94} A.L.~Melott, T.F.~Pellman, and S.F.~Shandarin,
  MNRAS {\bf 269}, 626 (1994).
\bibitem{cweb} J.R.~Bond, L.~Kofman, and D.~Pogosyan, Nature {\bf
    380}, 603 (1996).  
\bibitem{perc1} Ya.B.~Zel'dovich, Sov. Astron. Lett. {\bf 8}, 102
  (1982).  
\bibitem{perc2} S.F.~Shandarin and Ya.B.~Zel'dovich, Comments
  Astrophys. {\bf 10}, 33 (1983).  
\bibitem{percstat} S.F.~Shandarin, Sov. Astron. Lett. {\bf 9}, 104
  (1983). 
\bibitem{sh-zel-84} S.F.~Shandarin and Ya.B.~Zeldovich, Phys. Rev.
  Lett. {\bf 52}, 1488 (1984). 
\bibitem{yess-sh-96} C.~Yess and S.F.~Shandarin, Astrophys.~J. {\bf
    465}, 2 (1996).
\bibitem{colombi} S.~Colombi, D.~Pogosyan, and T.~Souradeep,
  Phys. Rev. Lett. {\bf 85}, 5515 (2000).
\bibitem{kly-sh-93} A.~Klypin and S.F.~Shandarin, Astrophys.~J. {\bf
    413}, 48 (1993). 
\bibitem{pm} K.~Heitmann, P.M.~Ricker, M.~Warren, and S.~Habib,
  Astrophys. J. Supp. {\bf 160}, 28 (2005). 
\bibitem{dor-zel-74} A.G.~Doroshkevich and Ya.B.~Zel'dovich,
  Astrophys. Space Sci. {\bf 35}, 55, (1974). 
\bibitem{percans} D.~Stauffer and A.~Aharony, {\em Introduction to
    Percolation Theory}, Taylor \& Francis, London (1992), Ch.~III and
  associated references. 
\bibitem{ziman} J.M.~Ziman, {\em Models of Disorder}, Cambridge
  University Press, Cambridge  (1979).
\bibitem{mon-yag-75} A.S.~Monin and A.M.~Yaglom, {\em Statistical
    Fluid Mechanics: Mechanics of Turbulence}, v. 2, The MIT Press,
  Cambridge U.S.A. and London U.K. (1975).  
\bibitem{sah-sat-sh-97} V.~Sahni, B.S.~Sathyaprakash, and
  S.F.~Shandarin,  Astrophys.~J. {\bf 476}, L1 (1997). 
\bibitem{euler-ch} A.G. Doroshkevich, Astrophysica {\bf 6}, 320
  (1970). 
\bibitem{bar-etal-86} J.M.~Bardeen, J.R.~Bond, N.~Kaiser, and
  A.S.~Szalay, Astrophys.~J. {\bf 304}, 16 (1986).
\bibitem{gott-etal-09} J.R.~Gott III, Y.Y. Choi, C. Park, and J. Kim,
  Astrophys. J. {\bf 695}, L45 (2009).
  \bibitem{z-space-anis} N. Kaiser, MNRAS {\bf 227}, 1 (1987); 
  S. Shandarin, JCAP {\bf 02}, 31 (2009)
\bibitem{perc-observ} Ya.B.~Zel'dovich, J.~Einasto, and S.F.~Shandarin,
  Nature {\bf 300}, 407 (1982); J.~Einasto, A.A.~Klypin, J.~Saar, and
  S.F.~Shandarin, MNRAS {\bf 206}, 529 (1984); S.F.~Shandarin and
  C.~Yess, Astrophys.~J. {\bf 505}, 12 (1998).

\end{thebibliography}
\end{document}